\documentclass[conference]{IEEEtran}

\usepackage[pdftex]{graphicx}
\usepackage{amsmath}
\usepackage{url}
\usepackage{amsfonts}
\usepackage{subfig}
\usepackage[linesnumbered,ruled]{algorithm2e}
\usepackage{verbatim}
\usepackage[noend]{algpseudocode}
\usepackage{xspace}
\usepackage{cite}

\usepackage[T1]{fontenc}
\usepackage{inconsolata}

\usepackage{amsmath,amsthm,amssymb,multirow,tabls,bigstrut}
\usepackage{tabularx}
\usepackage[table,xcdraw]{xcolor}
\usepackage{diagbox}

\usepackage{color}

\newcommand{\showComments}{yes}

\newtheorem{definition}{\bf Definition}

\def\squareforqed{\hbox{\rlap{$\sqcap$}$\sqcup$}}
\def\qed{\ifmmode\squareforqed\else{\unskip\nobreak\hfil
\penalty50\hskip1em\null\nobreak\hfil\squareforqed
\parfillskip=0pt\finalhyphendemerits=0\endgraf}\fi}

\definecolor{placeholderbg}{rgb}{0.85,0.85,0.85}

\newcommand{\note}[2]{
    \ifthenelse{\equal{\showComments}{yes}}{\textcolor{#1}{#2}}{}
}

\title{LRC: Dependency-Aware Cache Management\\for Data Analytics Clusters}
\author{
\IEEEauthorblockN{Yinghao Yu, Wei Wang, Jun Zhang, Khaled Ben Letaief\\}
\IEEEauthorblockA{Hong Kong University of Science and Technology\\
\tt yyuau@connect.ust.hk, weiwa@cse.ust.hk \{eejzhang, eekhaled\}@ust.hk
}
}

\graphicspath{{figures/}}
\usepackage{microtype}  

\begin{document}

\maketitle

\begin{abstract}

Memory caches are being aggressively used in today's data-parallel systems such as Spark, Tez, and Piccolo. However, prevalent systems employ rather simple cache management policies---notably the Least Recently Used (LRU) policy---that are \emph{oblivious} to the application semantics of data dependency, expressed as a directed acyclic graph (DAG). Without this knowledge, memory caching can at best be performed by ``guessing'' the future data access patterns based on historical information (e.g., the access recency and/or frequency), which frequently results in inefficient, erroneous caching with low hit ratio and a long response time.

In this paper, we propose a novel cache replacement policy, Least Reference Count (LRC), which exploits the application-specific DAG information to optimize the cache management. LRC evicts the cached data blocks whose \emph{reference count} is the smallest. The reference count is defined, for each data block, as the number of dependent child blocks that have not been computed yet. We demonstrate the efficacy of LRC through both empirical analysis and cluster deployments against popular benchmarking workloads. Our Spark implementation shows
that, compared with LRU, LRC speeds up typical applications by $60\%$.

\end{abstract}

\section{Introduction}
\label{sec:intro}


Data analytics clusters are shifting from on-disk processing toward in-memory computing. The increasing demand for interactive, iterative data analytics and the stalling speed of disk I/O force data-parallel systems to persist large volumes of data in memory to provide low latency \cite{power2010piccolo,zaharia2012resilient,ananthanarayanan2012pacman:,li2014tachyon}. Despite the continuous drop in RAM prices and the increasing availability of high-RAM servers, memory cache remains a constrained resource in large clusters. Efficient cache management, therefore, plays a pivotal role for in-memory data analytics.

\begin{figure}[htbp]
  \centering\includegraphics[width=4cm]{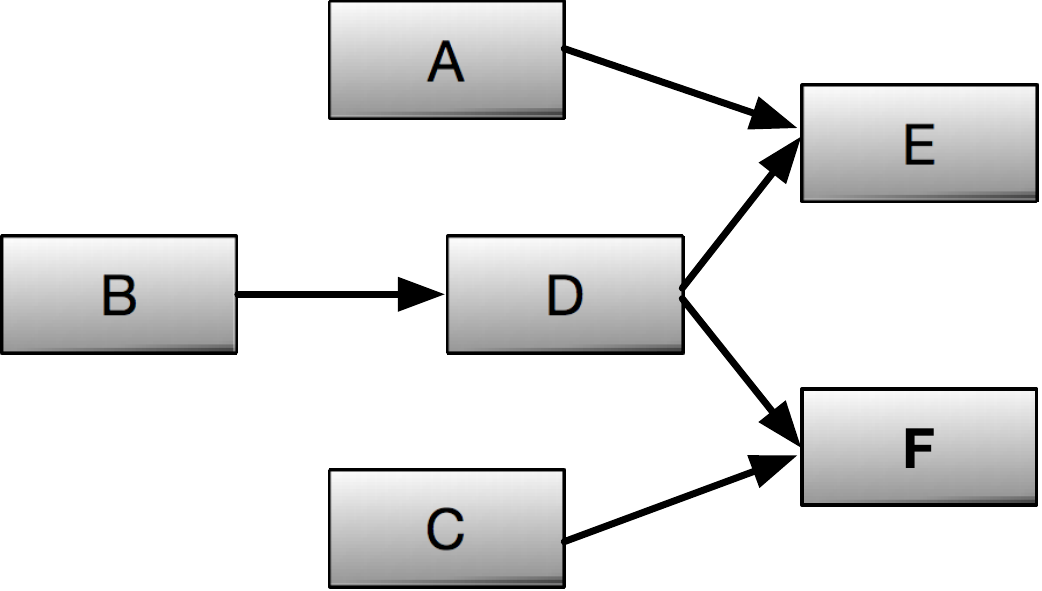}
  \caption{An example of the data dependency DAG of an application. Each block represents a dataset. Blocks $A$, $B$, and $C$ are input datasets already cached in memory. Block $D$ is an intermediate dataset derived from $B$. Block $E$ ($F$) is the final result derived from both $A$ and $D$ ($C$ and $D$).}
  \label{DAGExample}
\end{figure}

Caching is a classical problem and has been well studied in storage systems, databases, operating systems, and web servers. Yet, caching in data-parallel clusters has a defining characteristic that differentiates it from previous systems: cluster applications have clear semantics of data dependency, expressed as directed acyclic graphs (DAGs) of compute tasks. The DAG sketches out the task execution plan which dictates the underlying data access pattern, i.e., how data is computed and reused as input of descendant tasks. Fig.~\ref{DAGExample} shows an example DAG, where the computations of data blocks $E$ and $F$ depend on block $D$, and the corresponding tasks can only be scheduled after block $D$ has been derived from block $B$.

However, existing caching policies in prevalent data analytic systems \cite{zaharia2012resilient,marz2014apachestorm,shinnar2012m3r:} are \emph{agnostic} to the data dependency DAGs. Instead, they predict application-specific data access patterns based on historical information (e.g., frequency and recency). For instance, in Spark, the \texttt{BlockManager} evicts data blocks that are \emph{least recently used} (LRU) in the presence of high memory pressure \cite{zaharia2012resilient}. The LRU policy can be highly inefficient with low hit ratio. Referring back to the example of Fig.~\ref{DAGExample}, suppose that the cache can hold 3 blocks (assuming unit block size), and blocks $A$, $B$ and $C$ are cached at the beginning. Now to derive block $D$ and persist it in memory, the LRU policy will evict either block $A$ or $C$, as block $B$ has been recently used in the computation of block $D$, even though it will not be referenced again in the future. 
We show through this toy example that simply relying on the historical information can result in erroneous cache decisions.

In this paper, we ask how the application semantics of data dependency should be exploited to optimize cache management? Ideally, the solution should take full use of the DAG information and can be easily implemented as an efficient cache management policy with low overhead. In addition, the policy should be generally applicable to a wide range of in-memory computing frameworks. 


We propose a novel cache replacement policy, Least Reference Count (LRC), that always evicts the data block whose \emph{reference count} is the smallest. The reference count is defined, for each data block, as the number of \emph{unmaterialized} child blocks derived from it.


LRC provides benefits over existing cache management policies in three aspects. First, LRC can timely detect \emph{inactive} data blocks with \emph{zero} reference count. Such blocks are likely not be used again in the remaining computations\footnote{Unless re-computation is needed due to machine failures or stragglers.} and can be safely evicted from the memory. 
Second, compared with the historical information such as the block access recency and frequency, the reference count serves as a more accurate indicator of the likelihood of future data access. Intuitively, the higher the reference count a block has, the more child blocks depend on it, and the more likely the block is needed in the downstream computations. We show through empirical studies in Sec.~\ref{sec:background} that caching data blocks with the highest reference count increases the hit ratio by up to 119\% in comparison with caching the most recently used data. Third, the reference count can be accurately tracked at runtime with negligible overhead, making LRC a lightweight solution for all DAG-based systems.


We have prototyped LRC as a pluggable memory manager in Spark (details in Sec.~\ref{sec:lrc}). To demonstrate the efficacy of LRC, we have conducted extensive evaluations through EC2 deployment, including both the single- and multi-tenant experiments at scale. Compared to LRU---the default cache management policy in Spark---LRC retains the same application performance using only 40\%
of cache spaces. When operating with the same memory footprint, LRC is capable of
reducing the application runtime by up to 60\%. Our implementation can be easily adapted to other DAG-based data parallel systems and can also be extended to multi-tenant cache sharing systems such as Tachyon \cite{li2014tachyon}.

The remainder of the paper is organized as follows. In Sec.~\ref{sec:background}, we demonstrate the inefficiency of existing cache policies by characterizing data access patterns in parallel processing systems through empirical studies. We present the design of LRC policy and elaborate its implementation details in Sec.~\ref{sec:lrc}. Evaluation results are reported in Sec.~\ref{sec:evaluation}. We survey related work in Sec.~\ref{sec:related} and conclude the paper in Sec.~\ref{sec:conclusion}.

\section{Background And Motivation}
\label{sec:background}

In this section, we give the background information and motivate the need for a new cache policy through empirical studies. Unless otherwise specified, we shall limit our discussion to the context of Spark \cite{zaharia2012resilient}. However, nothing precludes applying the discussion to other frameworks such as Tez \cite{saha2015apache} and Storm \cite{marz2014apachestorm}.

\subsection{Semantics of Data Dependency}

Cluster applications such as machine learning, web search, and social network typically consist of complex \emph{workflows}, available as directed acyclic graphs (DAGs) to the cluster scheduler. These DAGs provide rich semantics of the underlying data access patterns, which entails a myriad of cache optimization opportunities that, so far, have not been well explored.

In Spark, data is managed through an easy-to-use memory abstraction called Resilient Distributed Datasets (RDDs) \cite{zaharia2012resilient}. An RDD is a collection of immutable datasets partitioned across a group of machines. Each machine stores a subset of RDD partitions (blocks), in memory or on disk. An RDD can be created directly from a file in a distributed storage system (e.g., HDFS \cite{shvachko2010hadoop}, Amazon S3 \cite{S3}, and Tachyon \cite{li2014tachyon}) or computed from other RDDs through a user-defined transformation. The DAG presents the workflow of RDD computations.



Whenever a job is submitted to the Spark driver, its DAG of RDDs becomes readily available to a driver component, \texttt{\texttt{DAGScheduler}} \cite{zaharia2012resilient}. The \texttt{DAGScheduler} then traverses the job DAG using depth-first search (DFS) and continuously submits \emph{runnable} tasks (i.e., those whose parent RDDs have all been computed) to the cluster scheduler to compute unmaterialized RDDs. Therefore, the cache manager can easily retrieve the DAG information from the \texttt{DAGScheduler}.
This information sheds light into the underlying data access patterns, based on which the cache manager can decide which RDD block should be kept in memory.

It is worth emphasizing that the availability of data dependency DAGs of compute jobs is not limited to Spark, but generally found in other parallel frameworks such as Apache Tez \cite{saha2015apache}: the Tez programming API allows the programmer to explicitly define the workflow DAG of an application, which is readily available to the Tez scheduler beforehand.

We caution that the actual data access sequence, though critically depends on
the dependency DAGs, cannot be fully characterized \emph{beforehand}. To see this, we refer back
to the previous example in Fig.~\ref{DAGExample}. After block $D$ has been computed, the cluster scheduler submits two task sets, say $T_1$ and $T_2$, to respectively compute data blocks $E$ and $F$ (noting that $A$ and $C$ are assumed to be in memory at the beginning). In this case, which block, $E$ or $F$, is computed first depends on the scheduling order of $T_1$ and $T_2$, which dictates the access order of blocks $A$ and $C$. We see from this simple example that there remains some uncertainty in the data access sequence, even with the dependency DAG available \emph{a priori}. It is such an uncertainty that rules out the use of the optimal offline cache policy, Belady's MIN \cite{belady1966study}, to maximize the cache hit ratio.



\subsection{Recency- and Frequency-Based Cache Management}

Despite the availability of the dependency DAGs, prevalent cache management
policies are agnostic toward this information. Instead, they
predict data access patterns based on the historical information, notably
\emph{recency} and \emph{frequency}.

\begin{itemize}
  \item \textbf{Least Recently Used (LRU)}: The LRU policy \cite{mattson1970evaluation} makes room for new data by evicting the cached blocks that have not been accessed for the longest period of time.
LRU is the \emph{de facto} cache management policy deployed in today's in-memory data analytics systems \cite{zaharia2012resilient,marz2014apachestorm,saha2015apache,li2014tachyon}. It predicts the 
access pattern based on the short-term data popularity, meaning the recently accessed data is assumed
to be likely used again in the near future. 


  \item \textbf{Least Frequently Used (LFU)}: The LFU policy \cite{aho1971principles} keeps
track of the access frequency of each
data block, and the one that has been accessed the least frequently 
has the highest priority to be evicted.
Unlike LRU, LFU predicts the access pattern based on the long-term data popularity, meaning
the frequently accessed data is assumed to be likely used again in the future.
\end{itemize}

Both LRU and LFU are very easy to implement. However, their obliviousness to
the application semantics of data dependency frequently result in inefficient,
even erroneous, cache decisions, as we show in the next subsection.

\subsection{Characterizing the Data Access Pattern}
\label{sec:data_access_patter}

To illustrate the need for being dependency-aware, we characterize the data
access patterns in typical analytics benchmarks through empirical studies. We
show that simply relying on the recency and frequency information for cache
management would waste a large portion of memory
to persist \emph{inactive data} that will never be used in downstream computations.

\vspace{.4em}
\noindent \textbf{Methodology.}
We ran SparkBench \cite{li2015sparkbench:}, a comprehensive benchmarking suite, in an Amazon EC2 \cite{ec2} cluster consisting of 10 \texttt{m4.large} instances. We measured the memory footprints and
characterized the data access patterns of 15 applications in SparkBench, including machine learning, graph computation, SQL queries, streaming, etc. Table~\ref{tab:SparkBench-all} summarizes the workload suite we used in our empirical studies.

\begin{table}[tbp]
  \centering
  \renewcommand{\arraystretch}{0.8}
  \caption{An overview of SparkBench suite \cite{li2015sparkbench:}.}
    \begin{tabularx}{.48\textwidth}{@{} *2{|>{\centering\arraybackslash}X} @{}|}
    \hline
    \textbf{Application Type } & \textbf{Workload } \bigstrut\\
    \hline
    \multirow{3}[6]{*}{Machine Learning } & \emph{Logistic Regression} \bigstrut\\
\cline{2-2}          & \emph{Support Vector Machine (SVM)} \bigstrut\\
\cline{2-2}          & \emph{Matrix Factorization} \bigstrut\\
    \hline
    \multirow{3}[6]{*}{Graph Computation } & \emph{Page Rank} \bigstrut\\
\cline{2-2}          & \emph{SVD Plus Plus} \bigstrut\\
\cline{2-2}          & \emph{Triangle Count} \bigstrut\\
    \hline
    \multirow{2}[4]{*}{SQL Queries} & \emph{Hive} \bigstrut\\
\cline{2-2}          & \emph{RDD Relation} \bigstrut\\
    \hline
    \multirow{2}[4]{*}{ Streaming Workloads } & \emph{Twitter Tag} \bigstrut\\
\cline{2-2}          & \emph{Page View} \bigstrut\\
    \hline
    \multirow{5}[10]{*}{Other Workloads } &  \emph{Connected Component}  \bigstrut\\
\cline{2-2}          & \emph{Strongly Connected Component} \bigstrut\\
\cline{2-2}          &  \emph{Shortest Paths} \bigstrut\\
\cline{2-2}          & \emph{Label Propagation} \bigstrut\\
\cline{2-2}          & \emph{Pregel Operation}  \bigstrut\\
    \hline
    \end{tabularx}%
  \label{tab:SparkBench-all}%
\end{table}%

\vspace{.4em}
\noindent \textbf{Data Access Patterns.}
Our experiments have identified two common access patterns across applications.

\begin{figure}[tbp]
  \centering\includegraphics[width=6cm]{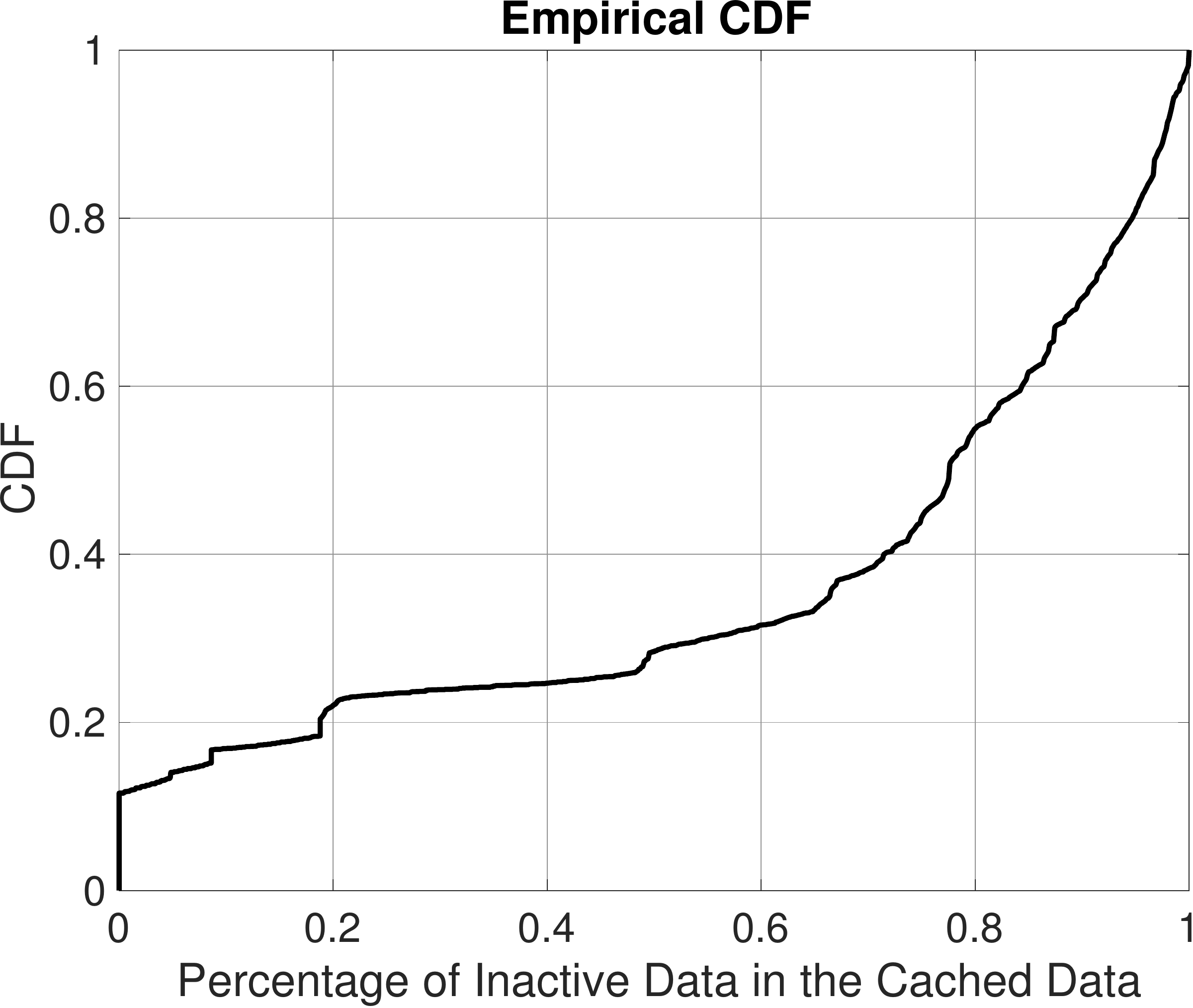}
  \caption{Distribution of inactive data cached during the execution of SparkBench \cite{li2015sparkbench:}.}
  \label{DeadDataCDF}
\end{figure}

\emph{1) Most data goes inactive quickly and will never be referenced again.}
  Fig.~\ref{DeadDataCDF} shows the distribution of inactive data cached in memory throughout the execution of 15 applications. We find that inactive data accounts for a large portion of memory footprint during the execution, with the median and 95th percentile being 77\% and 99\%, respectively. The dominance of inactive data blocks is in line with the data flow model where intermediate data is used by ``nearby'' computations in DAGs and therefore has a short life cycle. Keeping inactive data long in memory wastes cache spaces, inevitably resulting in low hit ratio.

\begin{figure}[tbp]
  \centering\includegraphics[width=6cm]{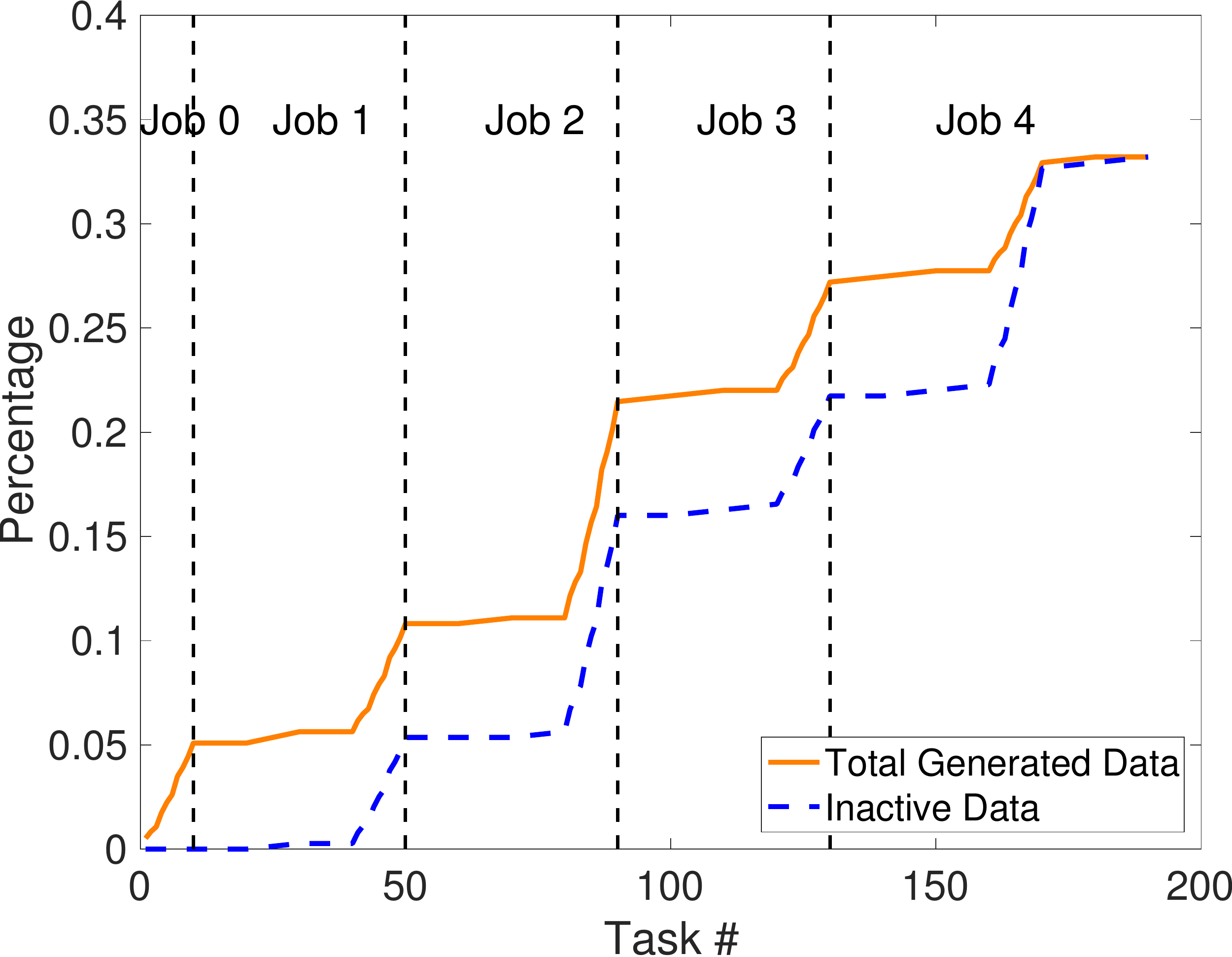}
  \caption{Memory footprints of intermediate data generated and inactive blocks cached during the execution of \emph{Connected Component}, an application in SparkBench \cite{li2015sparkbench:}.}
  \label{JobDeadData}
\end{figure}

\emph{2) Data goes inactive in \emph{waves}, often aligned with the generation of new data.} We further micro-benchmarked the memory footprint of total generated data against that of inactive data blocks cached during the execution of each application. Fig.~\ref{JobDeadData} shows the results for a representative application that computes the connected components of a given graph. The execution progress is measured in terms of the number of tasks completed. We see clearly that the data is produced and consumed in \emph{waves}, shown in Fig.~\ref{JobDeadData} as the lockstep along with the submission of new jobs following the DAG. Meaning that when a child RDD has been computed, its parents may go inactive. 




\begin{figure}[tb]
  \centering\includegraphics[width=7cm]{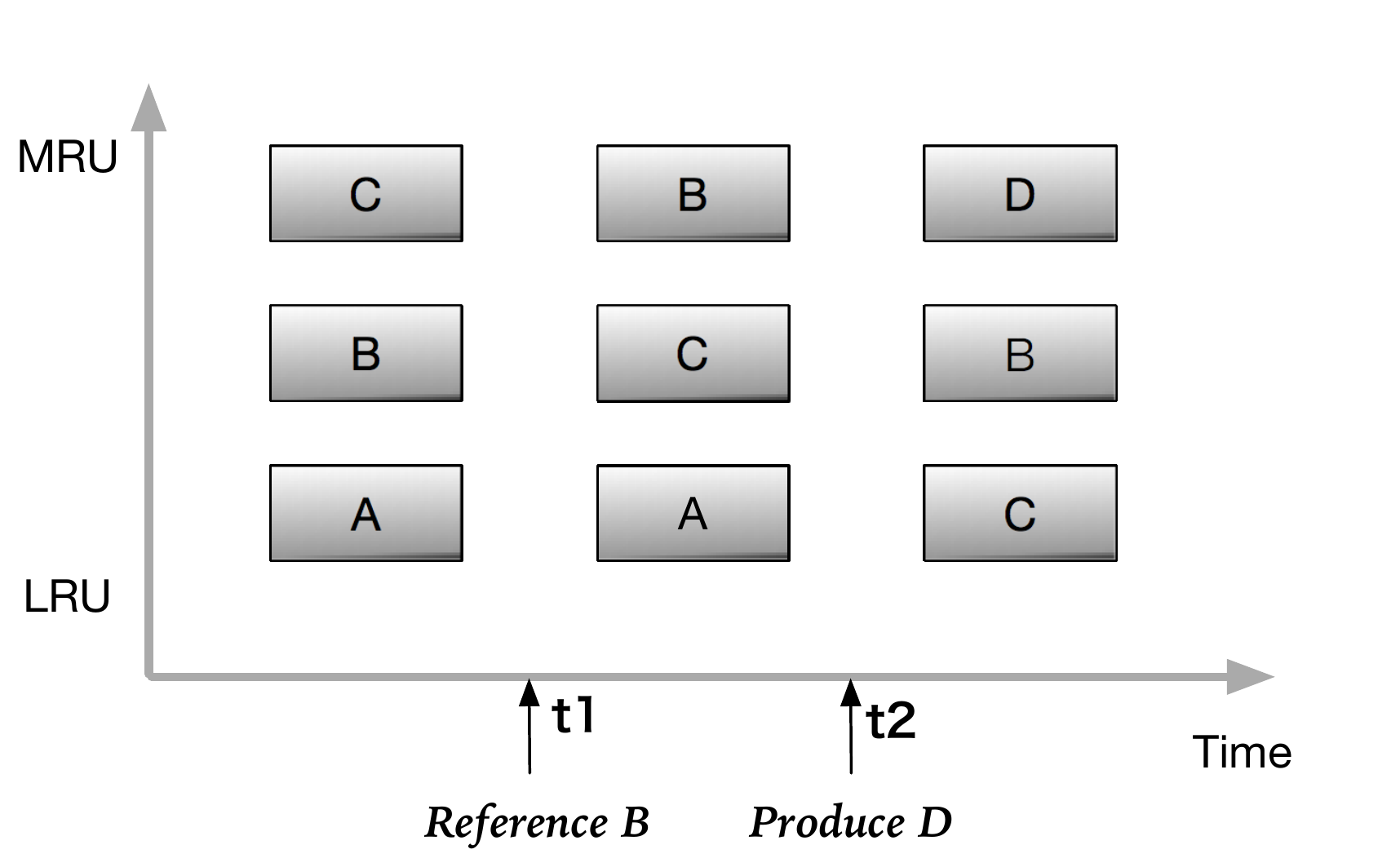}
  \caption{An example showing that LRU is unable to timely detect and evict
  inactive data.}
  \label{LRUCache}
\end{figure}

\vspace{.4em}
\noindent \textbf{Inefficiency of LRU and LFU.}
We learn from the empirical studies that the key to efficient cache
replacement is to timely evict inactive data blocks. Unfortunately, neither
LRU nor LFU is capable of doing so. We refer back to the example of
Fig.~\ref{DAGExample}, where each block is of a unit size, and the memory
cache  can persist three blocks in total. We start with LRU. Without loss
of generality, assume that the three blocks $A$, $B$, and $C$ are already in
memory at the beginning, with the recency rank $C > B > A$, i.e., from the most-%
recently-used (MRU) position to the least-recently-used (LRU).
Fig.~\ref{LRUCache} illustrates what happens in an LRU cache when data block
$D$ is materialized and then cached. Since block $D$ is derived from $B$, the
latter is firstly referenced as an input at time $t_1$ and is elevated to the
MRU position.  Soon later, block $D$ has been materialized at time $t_2$ and
is cached at the  MRU position, pushing the least-recently-used block $A$ out
of the memory. However, this would incur expensive tear-and-wear cost, in that
block $A$ will soon be reloaded in memory  to compute block $E$. In fact, we
see that the optimal decision is to evict block $B$, as it becomes inactive
and will never be used again. This simple example shows that
LRU is unable to evict inactive data in time, but it has to
wait \emph{passively} until the data demotes to the LRU position, which may
take a long time. 

We note that the LFU policy suffers from a similar problem. In the
previous example, to cache block $D$, the LFU policy would evict either block
$A$ or $C$ while retaining block $B$ in memory, because block $B$ has a
historical access record (in the computation of $D$), but the other two do
not.

To summarize, simply relying on the historical information, be it access
recency or frequency, is incapable of detecting inactive data in time. Efficient cache
management therefore must factor in the semantics of data dependency DAG. We
show how this can be achieved in the next section.

\section{Dependency-Aware Cache Management}
\label{sec:lrc}

In this section, we present a new cache management policy, Least Reference
Count (LRC), which makes the cache replacement decisions based on the data
dependency DAG. We also describe our implementation in Spark.

\subsection{Least Reference Count (LRC)}

We start with the definition of \emph{reference count}.

\begin{definition}[Reference count]
  For each data block $b$, the \emph{reference count} is define as the number of
  child blocks that are derived from $b$, but have not yet been computed.
\end{definition}

Back to the example of Fig.~\ref{DAGExample}, upon the DAG submission, blocks $A$, $B$, and $C$ 
all have reference count 1, while block $D$ has reference count 2.

The Least Reference Count (LRC) policy keeps track of the \emph{reference
count} of each data block, and whenever needed, it evicts the data
block whose \emph{reference count} is the smallest.

\vspace{.4em}
\noindent \textbf{Properties of LRC.}
The LRC policy has two nice properties that makes it highly efficient.

First, inactive data with zero reference count can be quickly detected and
evicted. Repeating the previous example of Fig.~\ref{DAGExample}, let blocks
$A$, $B$, and $C$ be in memory at the beginning, and the cache is full. Once
block $D$ has been computed, block $B$ becomes inactive with zero reference
count. Block $B$ is hence evicted to make room for block $D$.


Second, compared to recency and frequency, reference count serves as a more
accurate indicator of the likelihood of future data access. Intuitively, the
higher the reference count a block has, the more compute tasks depend on it,
and the more likely the block is needed in the near future. 

\begin{figure}[tb]
  \centering\includegraphics[width=6cm]{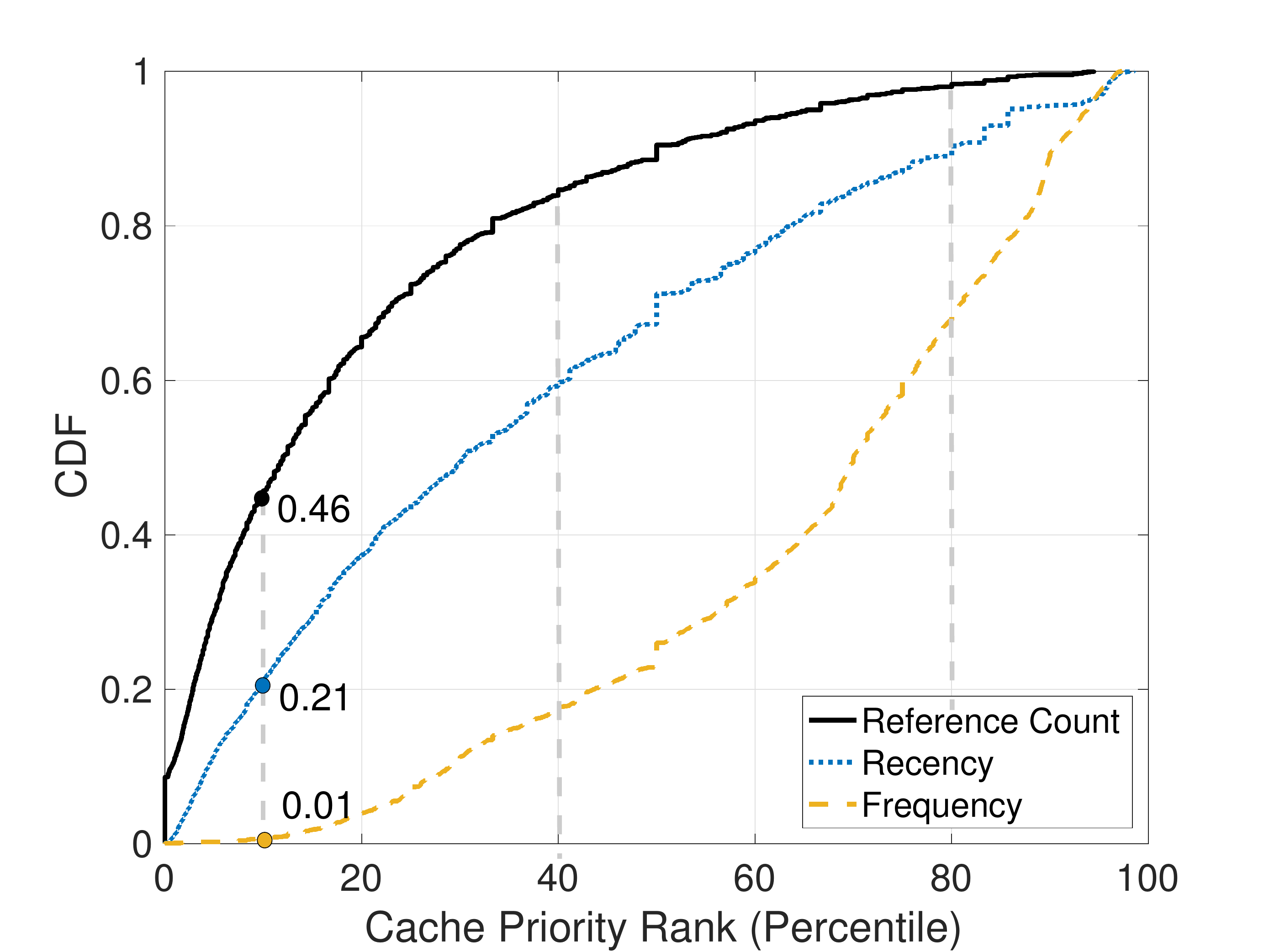}
  \caption{Distribution of the cache priority ranks of accessed data blocks with respect to
recency, frequency, and reference count. The workloads cover all SparkBench
applications \cite{li2015sparkbench:}.}
  \label{AccessCDF}
\end{figure}

To validate this intuition, we ran SparkBench applications on 10
\texttt{m4.large} EC2 instances. Specifically, whenever a data block is
accessed, we measure the \emph{cache priority rank} (in percentile) of the
block with respect to three metrics: the recency of last access, historical
access frequency, and reference count. A block having a top rank with respect
to a certain metric (say, top 1\% in recency) is likely cached in memory if
the corresponding cache policy is used (say, LRU). Fig.~\ref{AccessCDF} shows
the CDF of the measured ranks across 15 SparkBench applications. We see that
the reference count consistently gives higher ranks than the other two
metrics, meaning that it is the most accurate indicator of which  data will be
accessed next. For example, suppose that the cache can accommodate the
top 10\% ranked data blocks, each of a uniform size. Caching data blocks ranked
top in the reference count leads to the highest cache hit ratio of 46\%, which
is $2.19\times$ ($46\times$) compared with the recency (frequency) rank.


\subsection{LRC-Online}

We note that accurately computing the reference count requires extracting the
\emph{entire} data dependency DAG in an application. However, in systems such
as Spark and Tez, only the DAG information of a \emph{compute job} is readily
available, upon the job submission. A cluster application typically runs as a
workflow of several jobs. The data dependency between these jobs is usually
runtime information not available \emph{a priori}, and can be very complex if
jobs are executed iteratively. We address this challenge through two approaches.

First, studies of production traces reveal that a large portion of cluster
workloads are \emph{recurring} applications \cite{ferguson2012jockey}, which are run
periodically when new data becomes available. For these applications, we can
learn their DAGs from previous runs.

\begin{figure}[tbp]
  \centering\includegraphics[width=4cm]{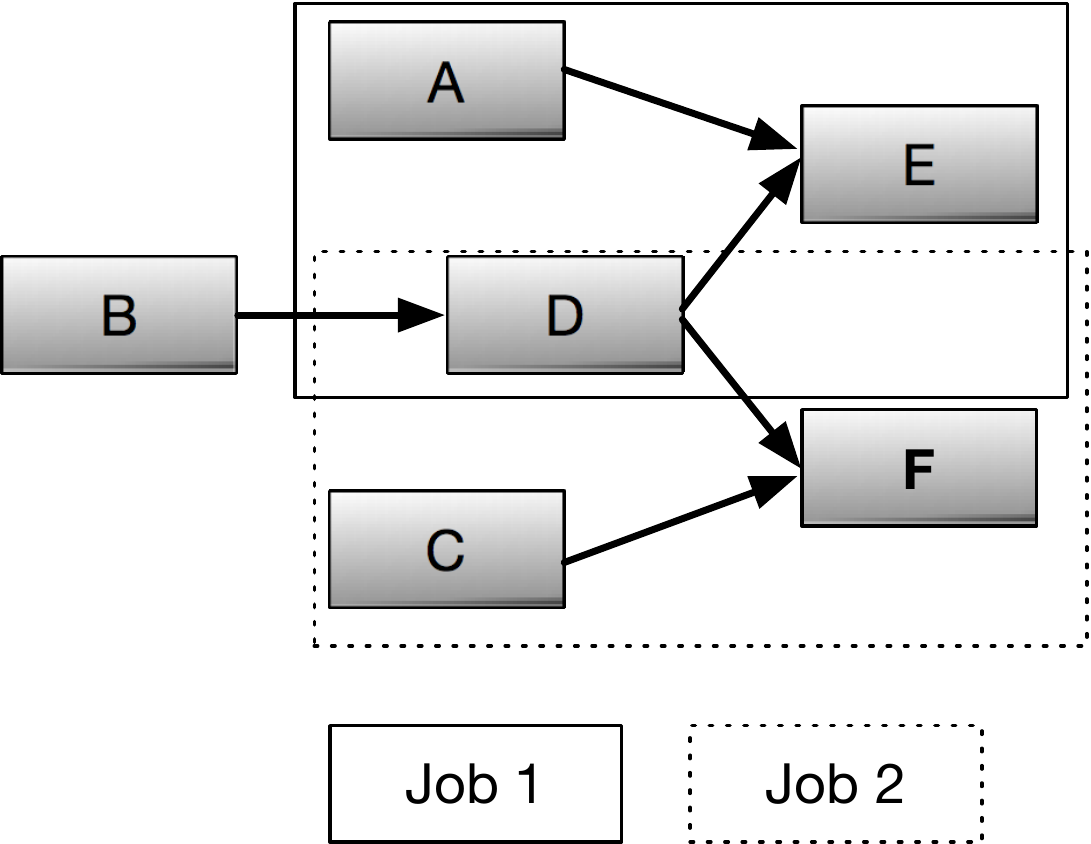}
  \caption{An example DAG where blocks $E$ and $F$ are respectively computed 
by Job 1 and Job 2. The former is scheduled earlier than the latter.}
  \label{DAGExampleByJob}
\end{figure}

Second, for the \emph{non-recurring} applications, such as interactive ad-hoc
queries, we propose \emph{LRC-Online} that online updates the data dependency
DAG of  the application whenever a new job is submitted.  Continuing the
example of Fig.~\ref{DAGExample}, we assume that the computations of block $E$
and $F$ belong to two different jobs, as illustrated in
Fig.~\ref{DAGExampleByJob}. Let Job 1 be submitted first, and the reference
count of block $D$ turns to 1. While this is inaccurate as block $D$ is also
used to compute block $F$, LRC-Online will soon correct its value after Job 2
has been submitted.


LRC-Online extends the application of LRC to the multi-tenant environments,
where multiple tenants share common input datasets. Despite the potential online
DAG profiling errors, our experiment results in Sec.~\ref{sec:evaluation} 
show that the performance of LRC-Online remains close to that of
LRC with the offline DAG information of applications.

\subsection{Spark Implementation}

We have prototyped LRC and LRC-Online as a pluggable block manager in Spark.
We next elaborate our implementation details.

\begin{figure}[tbp]
\centering\includegraphics[width=7cm]{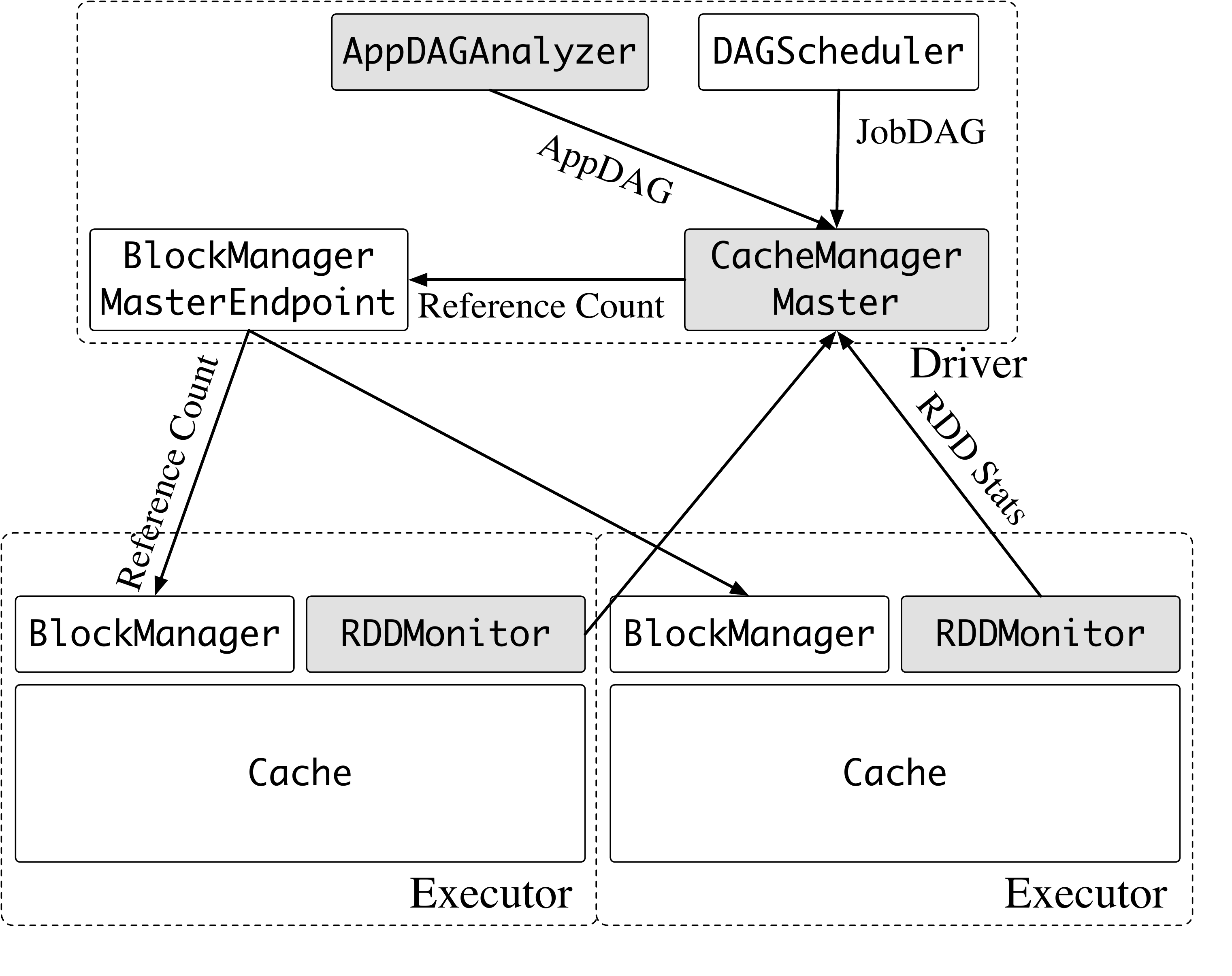}
\caption{Overall system architecture of the proposed application-aware cache manager in Spark. Our implementation modules are highlighted as shaded boxes.
}\label{Implementation}
\end{figure}

\vspace{.4em}
\noindent \textbf{Architecture overview.}
Fig.~\ref{Implementation} gives an architecture overview of our cache manager, where
the shaded boxes highlight our implementation modules. The cache manager consists of a
centralized controller on the master node and several distributed
\texttt{RDDMonitor}s that collect cache statistics on worker nodes and report
to the controller periodically. The controller has two key components: (1)
\texttt{AppDAGAnalyzer} that learns the data-dependency DAGs from previous runs for
recurring applications, or
in an online fashion for non-recurring applications, and (2) \texttt{CacheManagerMaster} that implements the main logic of the LRC and LRC-Online policies.
We summarize the key APIs of our implementation in Table~\ref{tab:API}.

\vspace{.4em}
\noindent \textbf{Workflow.}
Whenever an application is submitted, the Spark driver is launched on the
master node. The driver creates a \texttt{SparkContext} within which two
controller modules, \texttt{CacheManagerMaster} and \texttt{AppDAGAnalyzer},
are instantiated. By default, the driver also instantiates
\texttt{DAGScheduler} to parse job DAGs and
\texttt{BlockManagerMasterEndpoint} to communicate the cache information with
the \texttt{BlockManager} of workers. The driver then informs worker nodes to
launch Spark executors, which results in the deployment of
\texttt{BlockManager} and \texttt{RDDMonitor} across the cluster. Once the
connection between the executor and the driver has been established,
\texttt{RDDMonitor} starts to report RDD block status to
\texttt{CacheManagerMaster}, who maintains the reference count profile based
on the DAG information provided by \texttt{DAGScheduler} and
\texttt{AppDAGAnalyzer}. When the profile needs update due to a job
submission or a block access at the worker nodes, \texttt{CacheManagerMaster} notifies
\texttt{BlockManagerMasterEndPoint} in the driver to send updated reference
profile to \texttt{BlockManager} on the corresponding workers. The
\texttt{BlockManager} on each worker makes the eviction decisions \emph{locally}
based on the reference count profile it receives, and reports the events of
block access and eviction to \texttt{RDDMonitor} who then forwards the update
to the \texttt{CacheManagerMaster}.

\begin{table}[tbp]
  \centering
  \caption{Key APIs of our Spark implementation.}
\begin{tabularx}{.48\textwidth}{|l|X|X|r|}
    \hline
    \textbf{API } & \textbf{Description} \bigstrut\\
    \hline
    \texttt{parseDAG} & The \texttt{CacheManagerMaster} parses the DAG information obtained from the \texttt{DAGScheduler} and returns the profiled reference count for each RDD. \bigstrut\\
    \hline
    \texttt{updateReferenceCount} & The \texttt{CacheManagerMaster}  sends the parsed reference count profile to the corresponding \texttt{RDDMonitor}. \bigstrut\\
    \hline
    \texttt{reportRDDStatus} & The \texttt{RDDMonitor} informs the \texttt{CacheManagerMaster} of the status of the cached RDD blocks.\bigstrut\\
    \hline
    \texttt{updateReferenceCount} & Upon receiving the reference count updating message, the \texttt{RDDMonitor} updates reference counts of the corresponding RDD blocks.\bigstrut\\
    \hline
    \texttt{getReferenceCount} & When caching an RDD block, the \texttt{BlockManager} searches the local profile to get its reference count. If not found, ask the \texttt{BlockManagerMasterEndPoint} for update.\bigstrut\\
    \hline
    \texttt{decrementReferenceCount} & When an RDD block is referenced, the \texttt{RDDMonitor} deducts the its reference count in the maintained profile.\bigstrut\\
    \hline
    \texttt{evictBlocks} & When the cache is full, the \texttt{BlockManager} evicts the data with the least reference count.\bigstrut\\
    \hline
    \end{tabularx}%
  \label{tab:API}%
\end{table}%





\subsection{Discussion}

\noindent \textbf{Communication overhead.} To reduce the communication
overhead, each worker maintains the reference count profile locally and
synchronizes with the controller through a minimum number of message
exchanges. The \texttt{CacheManagerMaster} sends reference count updates to
the corresponding workers only when necessary. In particular, there are two
cases where the update is needed: 1) When a new job DAG is received from the
\texttt{DAGScheduler}, the \texttt{CacheManagerMaster} notifies workers to update
the reference count of the corresponding RDD blocks; 2) when an RDD block has
been referenced, and the block has replicas on the other workers, all those
workers should be notified to have a consistent reference count of the block. 
By default, RDD blocks are not replicated across the cluster, and our implementation
checks the Spark configuration to decide if such an update can be avoided.

\vspace{.4em} 
\noindent \textbf{Fault tolerance.} It is possible that a worker
may lose connection  to the driver at runtime, which results in a task
failure. In this case, the reference count profile maintained  by the
\texttt{CacheManagerMaster} will be inaccurate as the failed tasks will be
rescheduled soon. To address this inconsistency issue, the
\texttt{CacheManagerMaster} records the job ID upon receiving a job DAG from
\texttt{DAGScheduler}. This way, the \texttt{CacheManagerMaster} can quickly
detect job re-computation if the same job ID has been spotted before.
The consistency check for the reference count can then be applied.




\section{Evaluation}
\label{sec:evaluation}

In this section, we evaluate the performance of our cache manager through EC2
deployment against typical application workloads in SparkBench suite
\cite{li2015sparkbench:}. We first investigate how being dependency-aware
helps speed up a \emph{single} application with  much shorter runtime, and how
such a benefit can be achieved even when the DAG information is profiled
online. We next evaluate the performance of LRC in a \emph{multi-tenant}
environment  where multiple applications run in a shared cluster. These
applications have data dependency in between and compete for the
memory caches against each other.

\vspace{.4em}
\noindent \textbf{Cluster deployment.}
Our implementation is based on Spark 1.6.1. In order to highlight the
performance difference of memory read-write and disk I/O, we disabled the OS 
page cache using memory buffer by triggering direct
disk I/O from/to the hard disk. We deployed a 20-node EC2 cluster for the single-tenant experiments
and increased the cluster size to 50 nodes for the multi-tenant experiments.
Each node we used in the EC2 deployment is an \texttt{m4.large} instance \cite{ec2},
with a dual-core 2.4 GHz Intel Xeon E5-2676 v3 (Haswell) processor and 8
GB memory.


\begin{table}[tbp]
  \centering
  \caption{Impact of memory caches on the application runtime. We compare 
  against the two extreme cases: caching all data in memory versus caching none.}
    \begin{tabularx}{.48\textwidth}{|l|X|X|X|r|}
    \hline
    \textbf{Workload} & \textbf{Cache All} & \textbf{Cache None} \bigstrut\\
    \hline
    \emph{Page Rank} & 56 s   & 552 s \bigstrut\\
    \hline
    \emph{Connected Component} & 34 s   & 72 s \bigstrut\\
    \hline
    \emph{Shortest Paths} & 36 s   & 78 s \bigstrut\\
    \hline
    \emph{K-Means} & 26 s   & 30 s \bigstrut\\
    \hline
    \emph{Pregel Operation} & 42 s   & 156 s \bigstrut\\
    \hline
    \emph{Strongly Connected Component} & 126 s  & 216 s \bigstrut\\
    \hline
    \emph{Label Propagation} & 34 s   & 37 s \bigstrut\\
    \hline
    \emph{SVD Plus Plus} & 55 s   & 120 s \bigstrut\\
    \hline
    \emph{Triangle Count} & 84 s   & 99 s \bigstrut\\
    \hline
    \emph{Support Vector Machine (SVM)}   & 72 s   & 138 s \bigstrut\\
    \hline
    \end{tabularx}%
  \label{tab:characterizaiton}%
\end{table}%

\subsection{Single-Tenant Experiment}

We start with a simple scenario where a single tenant runs an application in a
small private cluster consisting of 20 nodes. We ran typical application
workloads in SparkBench and measured the cache hit ratio as well as the
application runtime using different memory management policies, including
LRU, LRC, and LRC-Online.


\vspace{.4em}
\noindent \textbf{Relevance of memory caches.}
It is worth emphasizing that memory caches may become irrelevant for some
applications. For example, we have observed in SparkBench that some applications 
are compute-intensive, and their runtime is mostly dictated by the CPU cycles. Some applications, 
on the other hand, need to shuffle large volumes of data,
and their performance is bottlenecked by the network. These applications benefit little from 
efficient cache management and do not see a significant runtime improvement even if
the system caches all data in memory.

In order to differentiate from these applications, we respectively measured
the application runtime in two extreme cases: 1) the system has large enough
memory and caches all data, and 2) the system caches no data at all. We
summarize our measurement results in Table~\ref{tab:characterizaiton}. We see
that some SparkBench applications, notably \emph{K-Means}, \emph{Label
Propagation} and \emph{Triangle Count}, have almost the same runtime in the
two cases, meaning that memory caches are irrelevant to their performance. We
therefore exclude these applications from evaluations 
but focus on four memory-intensive workloads
whose performance critically depends on cache management: \emph{Page Rank},
\emph{Pregel Operation}, \emph{Connected Component}, and \emph{SVD Plus Plus}.
Table~\ref{tab:testedworkloads} summarizes the input data sizes of these
workloads.

\begin{table}[tbp]
  \centering
  \caption{Summary of workload input data size.}
    \begin{tabularx}{.48\textwidth}{@{} *2{|>{\centering\arraybackslash}X} @{}|}
    \hline
    \textbf{Workload } & \textbf{Input Data Size} \bigstrut\\
    \hline
    \emph{Page Rank} & 480 MB \bigstrut\\
    \hline
    \emph{Pregel Operation} & 120 MB \bigstrut\\
    \hline
    \emph{Connected Component} & 280 MB \bigstrut\\
    \hline
    \emph{SVD Plus Plus} & 540 MB \bigstrut\\
    \hline
    \end{tabularx}%
  \label{tab:testedworkloads}%
\end{table}%
\begin{figure*}[tbp]
  \centering\includegraphics[width=1.0\textwidth]{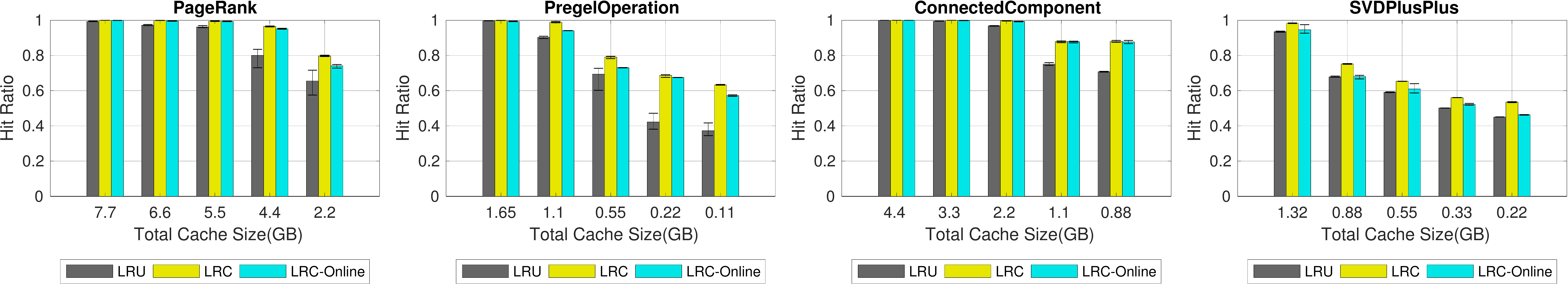}
  \caption{Cache hit ratio under the three cache management policies with different
  cache sizes.}
  \label{hitratio_single}
\end{figure*}

\begin{figure*}[tbp]
  \centering\includegraphics[width=1.0\textwidth]{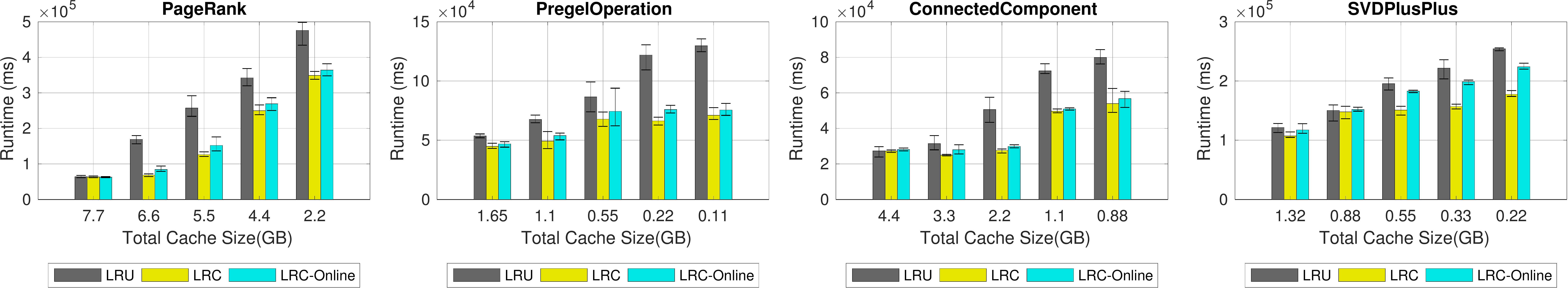}
  \caption{Application runtime under the three cache management policies with different
  cache sizes.}
  \label{runtime_single}
\end{figure*}

\vspace{.4em}
\noindent \textbf{Cache hit ratio and application runtime.}
We ran each application using
three cache replacement policies, i.e., LRU, LRC, and LRC-Online, with different memory cache sizes.
In particular, we configured \texttt{storage.memoryFraction} in the
legacy Spark to throttle the memory
used for RDD caching to a given size. We measured the cache hit ratio and the
application runtime against different cache sizes and depict the results in
Fig.~\ref{hitratio_single} and Fig.~\ref{runtime_single}, respectively.
The results have been averaged over 5 runs.

As expected, the less availability of the memory caches in the cluster, the smaller the cache 
hit ratio (Fig.~\ref{hitratio_single}), and the longer the application runs
(Fig.~\ref{runtime_single}). Regardless of the cache size, the two LRC algorithms
consistently outperform LRU, the default cache management policy in Spark, across all
applications. The benefits achieved by
the LRC policy, in terms of the application speedup, varies with different 
cache sizes as well as the application workloads. In particular, 
compared to the default LRU policy, our LRC algorithm reduces the
runtime of \emph{Page Rank} by 60\% (from 170 s to 64 s) when the cluster cache size is 5.5 GB. 
Table~\ref{tab:maximum saving} summarizes the largest runtime savings of LRC over LRU
for each application we evaluated. 

The efficiency of LRC policy can also be illustrated from a different
perspective,  in that LRC requires much smaller cache spaces than that of LRU,
but it is able to achieve the same  cache hit ratio. For example, to achieve
the target hit ratio of 0.7 for \emph{Pregel Operation}, LRU requires 0.55 GB
memory caches. In comparison, LRC requires only 0.22 GB, an equivalent of 60\% saving
of cache spaces.

We make another interesting observation that for different applications, the cache
hit ratio has different impact on their runtime. For example, the workload of
\emph{Page Rank} suffers from the most significant slowdown, from 67 s to 320
s, when the cache hit ratio decreases from 1 to 0.85. The reason is that the
computation of \emph{Page Rank} consists of some large RDDs, and their cache
miss  critically increases the total runtime. For \emph{Connected Component},
salient slowdown is observed when the cache hit ratio drops from 0.7 to 0.4;
for \emph{SVD Plus Plus}, we observe a linear slowdown with respect to the
decrease of cache hit ratio.

\begin{table*}[tbp]
  \centering
  \caption{Summary of the maximum reduction of application runtime over LRU.}
    \begin{tabular}{|c|c|c|c|c||c|c|}
    \hline
    \textbf{Workload} & \textbf{Cache Size} & \textbf{LRU}   & \textbf{LRC}   & \textbf{LRC-Online} & \textbf{Speedup by LRC} & \textbf{Speedup by LRC-Online} \bigstrut\\
    \hline
    \emph{Page Rank} & 6.6 GB & 169.3 s & 68.4 s & 84.5 s & 59.58\% & 50.06\% \bigstrut\\
    \hline
    \emph{Pregel Operation} & 0.22 GB & 121.9 s & 66.3 s & 75.9 s & 45.64\% & 37.74\% \bigstrut\\
    \hline
    \emph{Connected Component} & 2.2 GB & 50.6 s & 27.6 s & 29.9 s & 45.47\% & 40.97\% \bigstrut\\
    \hline
    \emph{SVD Plus Plus} & 0.88 GB & 254.3 s & 177.6 s & 223.9 s & 30.17\% & 11.96\% \bigstrut\\
    \hline
    \end{tabular}%
  \label{tab:maximum saving}%
\end{table*}%

\vspace{.4em}
\noindent \textbf{LRC-Online.}
As discussed in the previous section, when the \emph{entire} application DAG
cannot be retrieved \emph{a priori}, we can profile the submitted job DAGs at
runtime using LRC-Online. We now evaluate how such an online approach
can approximate the LRC policy with offline knowledge of application DAG. 
We see through Fig.~\ref{hitratio_single} and Fig.~\ref{runtime_single}, 
that LRC-Online is a close approximation of LRC for all applications
except \emph{SVD Plus Plus}. As illustrated in Table~\ref{tab:maximum
saving}, with online profiling, LRC can only speed up the application by
$12\%$, as opposed to $30\%$ provided by the offline algorithm.


We attribute the performance loss of LRC-Online to the fact that the
datasets generated in the current job might be referenced by another in the
future, whose DAG is yet available to the \emph{DAGScheduler}. Therefore, the
reference count of the dataset calculated at the current stage may not be
accurate. To quantify the inaccuracy of online profiling, we measure
\emph{reference distance}, for each data reference, as the number of
intermediate jobs from the \emph{source job} where the data is generated to
the \emph{destination job} where the data is used. Intuitively, the longer the
reference distance is, the greater chance it is that referencing the block in
the future may encounter a cache miss. This is because  without knowing the
entire application DAG beforehand,  LRC-Online can only tell the data
dependency in the current job, and will likely evict all the generated data
blocks after the source job has completed.

Table~\ref{tab:reference distance} summarizes the average reference distance of
the data blocks generated in each application. All applications but \emph{SVD Plus Plus}
have reference distance less than 1, meaning that most of the generated data is
likely used either by the current job or the next one. This result is in line with 
the observation made in Fig.~\ref{JobDeadData}, where intermediate data 
goes inactive in waves and in lockstep with job submission. 
\emph{SVD Plus Plus}, on the other hand, has the longest reference distance, 
which explains its performance loss with LRC-Online.

\begin{table}[htbp]
  \centering
  \caption{Average reference distance.}
    \begin{tabularx}{.48\textwidth}{@{} *2{|>{\centering\arraybackslash}X} @{}|}
    \hline
    \textbf{Workload} & \textbf{Average Reference Distance} \bigstrut\\
    \hline
    PageRank & 0.95 \bigstrut\\
    \hline
    PregelOperation & 0.73 \bigstrut\\
    \hline
    ConnectedComponent & 0.74 \bigstrut\\
    \hline
    SVDPlusPlus & 1.71 \bigstrut\\
    \hline
    \end{tabularx}%
  \label{tab:reference distance}%
\end{table}%

\begin{table}[htbp]
  \centering
  \caption{Summary of workloads used in the multi-tenant experiment.}
    \begin{tabularx}{.48\textwidth}{@{} *3{|>{\centering\arraybackslash}X} @{}|}
    \hline
    \textbf{Tenant Index} & \textbf{Workload} &  \textbf{Input Data Size}\bigstrut\\
    \hline
    1-8 & ConnectedComponent & 745.4 MB \bigstrut\\
    \hline
    9-16 & PregelOperation & 66.9 MB \bigstrut\\
    \hline
    \end{tabularx}%
  \label{tab:testedworkloads with multi-tenants}%
\end{table}%

To summarize, LRC-Online is a practical solution that well approximates
LRC and consistently outperforms the LRU policy across applications.

\subsection{Multi-Tenant Experiment}

We now investigate how our cache manager performs in a multi-tenant
environment through a 50-node EC2 deployment. Notice that the Spark driver
unifies the indexing of jobs and RDDs for all tenants. Once a job is
submitted, the RDDs in its DAG are indexed incrementally based on the unified
index. In this case, the offline RDD reference count is unavailable because
the order of actual job submission sequence from multiple tenants is uncertain
due to runtime dynamics. Therefore, the data dependency DAG can only be determined
online. For this reason, we compare the performance of LRC-Online against LRU.

In the first set of experiments, we emulated 16 tenants submitting jobs
simultaneously to the Spark driver, with different cache sizes. The
workload profile of the tenants is given in Table~\ref{tab:testedworkloads
with multi-tenants}.  In the second experiment, we fixed the total cache size
to be 1.26 GB and increased the number of tenants from 8 to 20.

\begin{figure}[htbp]
\centering\includegraphics[width=0.48\textwidth]{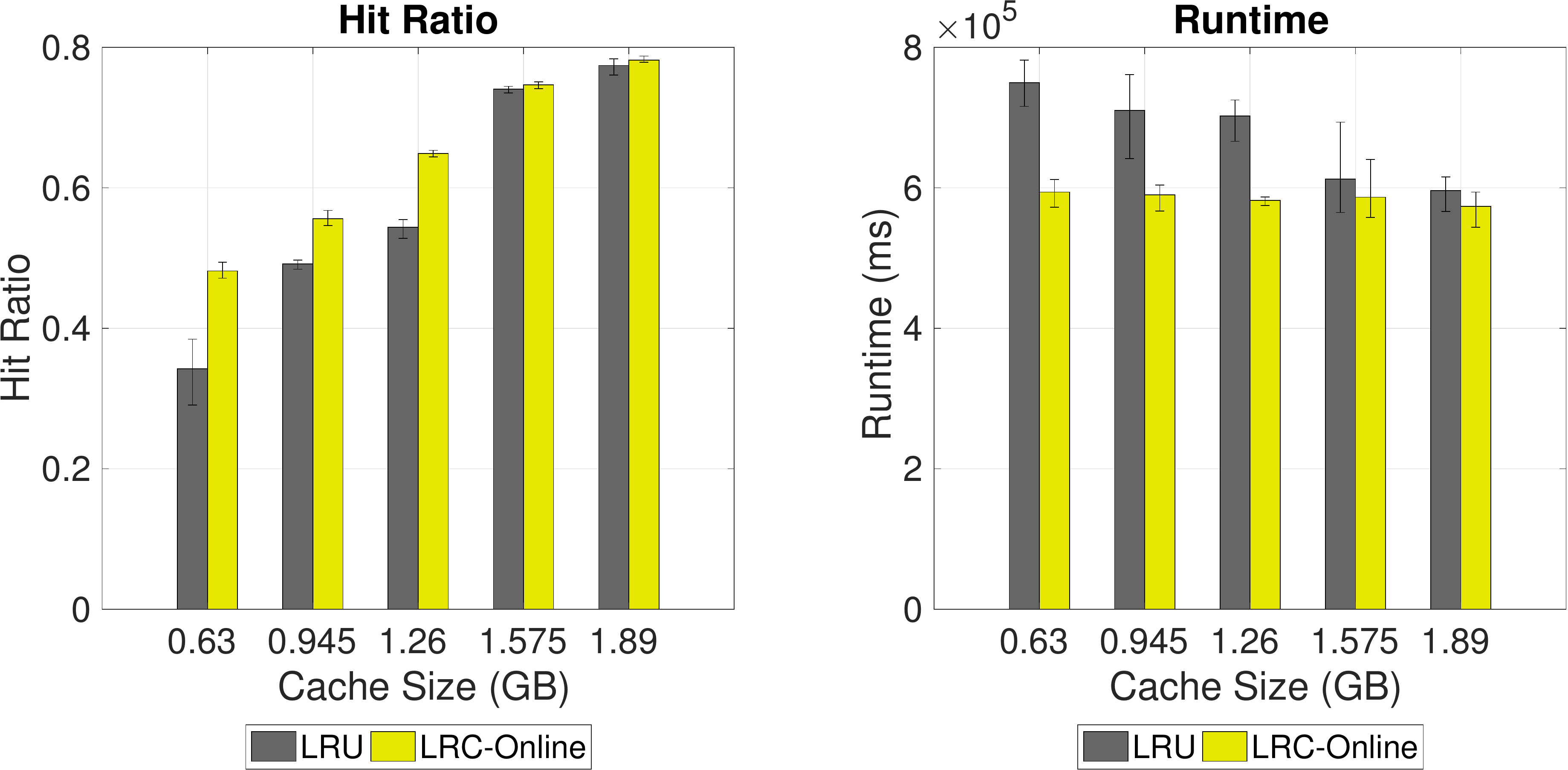}
\caption{Cache hit ratio and total runtime with different cache sizes in the multi-tenant experiment.}\label{MacroExp_Cache}
\end{figure}

\begin{figure}[htbp]
\centering\includegraphics[width=0.48\textwidth]{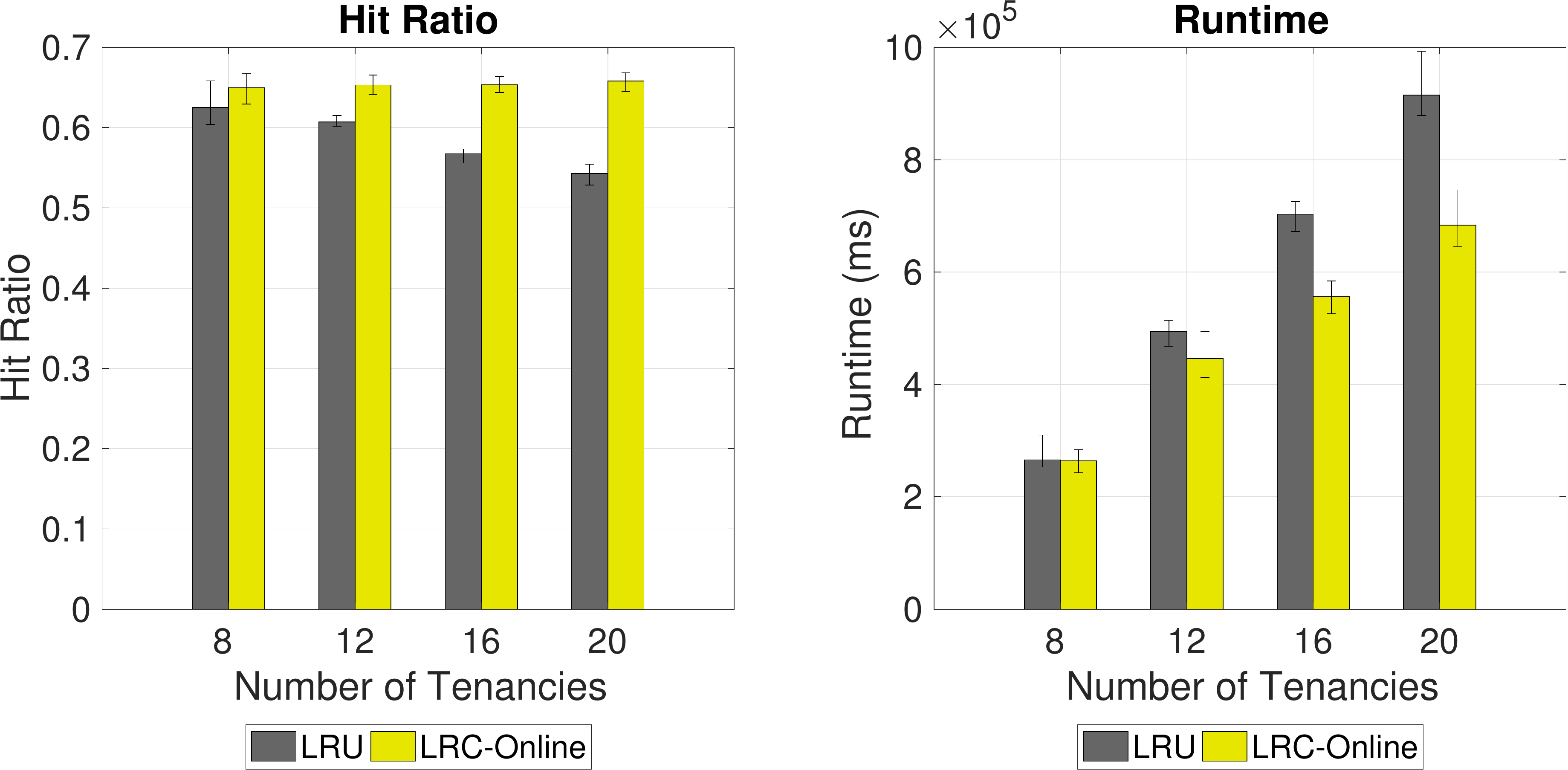}
\caption{Cache hit ratio and total runtime with different numbers of tenants.}\label{MacroExp_Tenant}
\end{figure}

Fig.~\ref{MacroExp_Cache} and Fig.~\ref{MacroExp_Tenant} show the results. We
find that LRC-Online is capable of achieving larger performance gains over LRU
with smaller cache sizes or with more tenants. This suggests that leveraging the
application semantics of data dependency is of great significance, 
especially when the cluster memory caches are heavily competed.

\section{Related Work}
\label{sec:related}

\noindent \textbf{Traditional Caching on a Single Machine.}
Memory caching has a long history and has been widely employed in storage systems \cite{chen1994raid}, databases \cite{stonebraker1981operating}, file systems \cite{nelson1988caching}, web servers \cite{cao1997cost}, operating systems \cite{baker2011megastore}, and processors \cite{henning2000spec}. Over the years, a vast amount of caching algorithms have been proposed. These algorithms run on a single machine and can be broadly divided into two categories:

\emph{Recency/frequency-based policies:} LRU \cite{mattson1970evaluation} and LFU \cite{aho1971principles} are the two widely used caching algorithms. As shown in Section \ref{sec:background}, neither algorithm adapts well to the data access pattern in data analytic clusters even though they are simple to implement. 

\emph{Hint-based policies:} Many cache policies evict/prefetch data blocks through hints from applications \cite{cao1996implementation,patterson1995informed}, which are provided by the programmers to indicate what data will be referenced again and when. Nevertheless, inserting such hints can be difficult to the programmer, who has to carefully examine the underlying data access pattern. 





\vspace{.4em}
\noindent \textbf{Cache Management in Parallel Processing Systems.}
Despite the significant performance impact of memory caches, cache management remains a relatively unchartered territory in data parallel systems. Prevalent parallel frameworks (e.g., Spark \cite{zaharia2012resilient}, Tez \cite{saha2015apache}, and Tachyon \cite{li2014tachyon}) simply employ LRU to manage cached data on cluster machines, which results in a significant performance loss \cite{ananthanarayanan2012pacman:,xu2016memtune:}.

To our knowledge, the recently proposed MemTune \cite{xu2016memtune:} is the only caching system that leverages the application semantics. MemTune dynamically adjusts the memory share for task computation and data caching in Spark and evicts/prefetches data as needed. As opposed to our proposal that accounts for the entire DAG and its dependencies, MemTune only considers locally dependent blocks of currently runnable tasks. Moreover, when it comes to a multi-tenant environment, the complexity of MemTune is also multiplied, as MemTune keeps track of all the submitted DAGs and traverses them at runtime to search for the downstream tasks whenever a task completes. In comparison, LRC parses the job DAG upon job submission and only sends the reference count profile to the corresponding worker node. The reference count updating message is light and simple compared to the DAG structure itself, and can be easily applied to any DAG-based system and even heterogeneous environments running different systems, like Tachyon \cite{li2014tachyon}. 

Recent works \cite{pu2016fairride:,kunjir2015robus} have also studied the problem of cache allocation for multiple users with shared files in clusters, where fairness is the main objective to achieve. We view these works orthogonal to our proposed research. Once the cache space of each application has been allocated by some fair sharing policies, we can leverage the application-specific semantics for efficient cache management.

\section{Concluding Remark}
\label{sec:conclusion}

In this paper, we proposed a dependency-aware cache management policy, Least
Reference Count (LRC), which evicts data blocks whose reference count is the
smallest. The reference count is defined, for each data block, as the number
of dependent child blocks that have not been computed yet. With LRC, inactive
data blocks can be timely detected and evicted, saving cache spaces for more
useful data. In addition, we showed that reference count serves as an
accurate indicator of the likelihood of future data access. 
We have implemented LRC as a pluggable cache manager
in Spark, and evaluated its performance through EC2 deployment. Experimental
results show that compared to the popular LRU policy, our LRC implementation 
is capable of achieving the same application
performance at the expense of only 40\% of cache spaces. When using the same
amount of memory caches, LRC can reduce the application runtime by up to 60\%. 


\section*{Acknowledgements}
\label{sec:ack}

We thank Chengliang Zhang for helping on the implementation of LRC-Online and the deployment of
multi-tenant experiments. We thank the anonymous reviewers for their invaluable feedback.
This research was partly supported by the Hong Kong Research Grants Council under Grant No.~16200214.

\bibliography{total}
\bibliographystyle{IEEEtran}

\end{document}